\documentclass[aps,prl,twocolumn,showpacs,amsmath,amssymb]{revtex4-1}
\usepackage{graphicx}
\usepackage{color}
\usepackage{soul}
\usepackage[colorlinks,plainpages=false,linkcolor=black,urlcolor=blue,citecolor=blue,pdfpagemode=UseNone,pdfstartview=FitH]{hyperref}

\newcommand{\be}{\begin{equation}}
\newcommand{\ee}{\end{equation}}

\newcommand{\bea}{\begin{eqnarray}}
\newcommand{\eea}{\end{eqnarray}}
\newcommand{\bd}{\begin{displaymath}}
\newcommand{\ed}{\end{displaymath}}
\newcommand{\ba}{\begin{array}}
\newcommand{\ea}{\end{array}}
\newcommand{\bi}{\begin{itemize}}
\newcommand{\ei}{\end{itemize}}
\newcommand{\bc}{\begin{center}}
\newcommand{\ec}{\end{center}}
\newcommand{\bfl}{\begin{flushleft}}
\newcommand{\efl}{\end{flushleft}}
\newcommand{\bfr}{\begin{flushright}}
\newcommand{\efr}{\end{flushright}}

\def\6{\partial}

\def\={\!\!\!&=&\!\!\!}
\def\+{\!\!\!&&\!\!\!+~}
\def\-{\!\!\!&&\!\!\!-~}

\newcommand \Tc{ $T_\mathrm{c}$}
\newcommand \Qaf{${\bf Q}_\mathrm{AF}$}

\begin{document}

\title[]{Incommensurate magnetic fluctuations and Fermi surface topology in LiFeAs}
%Goldstone mode and resonant magnetic excitation in the coexistence phase of iron-based superconductors}
%
\author {J. Knolle$^{1}$}
\email{jknolle@pks.mpg.de}
\author{V.B. Zabolotnyy$^{2}$}
\author{I. Eremin$^{3}$}
\email{ieremin@tp3.rub.de}
\author{S.V. Borisenko$^{2}$}
\author{N. Qureshi$^4$}
\author{M. Braden$^4$}
\author{D.V. Evtushinsky$^2$}
\author{T.K. Kim$^{2,5}$}
\author{A.A. Kordyuk$^2$}
\author{S. Sykora$^2$}
\author{Ch. Hess$^{2}$}
\author{I.V. Morozov$^{2,6}$}
\author{S. Wurmehl$^2$}
\author{R. Moessner$^{1}$}
\author{B. B\"uchner$^{2,7}$}
 \affiliation{$^1$Max Planck Institute for the Physics of Complex Systems, D-01187 Dresden, Germany}
  \affiliation{$^2$Leibniz-Institut f\"ur Festk\"orper- und Werkstoffforschung Dresden, P.O. Box 270116, D-01171 Dresden,Germany}
 \affiliation {$^3$Institut f\"ur Theoretische Physik III, Ruhr-Universit\"at Bochum, D-44801 Bochum, Germany}
 \affiliation {$^4$II. Physikalisches Institut, Universit\"at zu K\"oln, Z\"ulpicher Str. 77, D-50937 K\"oln, Germany}
% \affiliation{$^5$ Institute of Physics, Kazan Federal University, 420008 Kazan, Russian Federation}
 \affiliation{$^5$ Diamond Light Source, Didcot, OX11 0DE, United Kingdom}
 \affiliation{$^6$Moscow State University, 119991 Moscow, Russia}
\affiliation{$^7$ Institut f\"ur Festk\"orperphysik, Technische Universit\"at Dresden, D-01171 Dresden, Germany}
\begin{abstract}
Using the angle-resolved photoemission spectroscopy (ARPES) data accumulated over the whole Brillouin zone (BZ)
in LiFeAs we analyze the itinerant component of the dynamic spin susceptibility in this system in the normal and
superconducting state. We identify the origin of the incommensurate magnetic inelastic neutron scattering (INS)
intensity as scattering between the electron pockets, centered around the $(\pi,\pi)$
point of the BZ and the large two-dimensional hole pocket, centered around the $\Gamma$-point of the BZ.
As the magnitude of the superconducting gap within the large hole pocket is relatively small and angle dependent, we interpret the INS data in the superconducting state as a renormalization of the particle-hole continuum rather than a true spin exciton. Our comparison indicates that the INS data can be reasonably well described by both the sign changing symmetry of the superconducting gap  between electron and hole pockets as well as sign preserving gap, depending on the assumptions made for the fermionic damping.
 \end{abstract}

\date{\today}


\maketitle

The relation between unconventional superconductivity and
magnetism is one of the most interesting topics in
condensed-matter physics. For example, in most of the iron-based superconductors superconductivity occurs in close vicinity to an antiferromagnetic (AF) state\cite{kamihara,korshunov_review,chubukov_review}. Moreover, superconductivity emerges when antiferromagnetic order in parent compounds is
suppressed, either by electron/hole doping or disorder. In addition, short-range AF spin excitations are still present in the normal state of the doped systems and also become resonant in the superconducting state
%below \comm{ \Tc}
at energies below twice the superconducting gap magnitude, $2\Delta_0$\cite{lumsden_review}. This resonant enhancement is believed to be a signature of a certain phase structure of the superconducting gap as the paramagnetic  spin response of the Bogolyubov quasiparticles at the antiferromagnetic wave vector \Qaf{}  is sensitive to the anomalous coherence factor $1-\frac{\Delta_{\bf k} \Delta_{\bf k+Q_{AF}}}{|\Delta_{\bf k}| |\Delta_{\bf k+Q_{AF}}|}$. Once the superconducting gap at parts of the Fermi surface, connected by \Qaf, changes sign, the spin response acquires an additional enhancement at $\Omega \leq 2 \Delta_0$, which is a hallmark of unconventional superconductivity. The observation of the spin resonance  in many iron-based superconductors provides strong evidence for the so-called $s^{+-}$-wave symmetry of the superconducting gap, where the gap structure changes sign between electron and hole pockets\cite{kuroki,mazin,chubukov}. Note that this does not exclude the gap on each pocket to have a strong angular variation and even accidental nodal lines, allowed by $A_{1g}$ symmetry \cite{chubukov_review}. The angular variation of the gap, measured in ARPES\cite{Symmetry}, is inconsistent with idealized lattice version of $s^{+-}$, but can be modeled by taking into account interaction effects.

While the behavior, described above, is observed in the majority of the iron-based superconductors, there are some notable exceptions. Perhaps the most interesting one is the stoichiometric LiFeAs, which superconducts at \Tc=17\,K without any doping\cite{wang,chu,morozov}. In addition, LiFeAs shows neither static AF ordering nor nesting between electron and hole bands at {\bf Q}$_{AF}$\cite{borisenko}. Several neutron scattering experiments were performed recently in LiFeAs\cite{taylor,wang_dai,qureshi}, including only one study\cite{qureshi} on superconducting single crystals, where magnetic intensity at an incommensurate momentum close to $(\pi,\pi)$ was observed. Its renormalization across $T_c$ was found to be too weak to draw a definite conclusion about the phase structure of the superconducting gap. Furthermore, some controversy on the phase structure of the order parameter arises in the analysis of the quasiparticle interference in the superconducting state of LiFeAs\cite{haenke,allan}.

In this paper we use the ARPES data for LiFeAs, 111-type pnictide superconductor, which is known to be free from surface effects\cite{Lankau} to strengthen a connection to the INS response. This is particularly important, given the controversy on the Fermi surface topology in this system.\cite{borisenko,putzke} We employ an effective tight-binding fit to the high-quality LiFeAs photoemission data in order to compute the spin response within random phase approximation (RPA). We believe this procedure is only possible at present in 111 systems as availability of the requisite data is most complete here. A comparison with INS data shows that the incommensurate magnetic scattering intensity arises due to scattering between the electron pockets, centered around the $(\pi,\pi)$ point of the BZ, and the large two-dimensional hole pocket, centered around the $\Gamma$-point of the BZ. We also find that the renormalization of the neutron intensity upon opening of the superconducting gap is relatively weak,  consistent with the INS experiments.

{\it ARPES and tight-binding fit.}
Owing to its ability to resolve both momentum and energy of the electronic states,
modern photoemission can be used to map out a complete low-energy electronic structure of a layered compound, like LiFeAs. From such a comprehensive data set  one may extract the dispersion of quasiparticles at any momentum. This can be used to calculate numerous properties like heat capacity, plasma frequency or the Hall coefficient\cite{Stockert, Evtushinsky}. However to make this possible quasiparticle dispersions have
to be conveniently parameterized.  One way to do this is via a tight-binding fit. Indeed, tight-binding models including up to 10 bands, have been developed to fit the LDA band structure of the iron-pnictide superconductors \cite{Eschrig, Graser}. However from a practical perspective it is more
favorable to use an effective tight-binding model, separately describing dispersions of each band that crosses the Fermi level\cite{Inosov}.
In case of the square lattice with a tetragonal symmetry the quasiparticle dispersion can be fit by the following formula:
\begin{equation}
\mathcal{E}(k_x, k_y) = \sum_{m, n=0}^{N-1} \alpha_{m, n} \phi_{m, n}(k_x, k_y)\mbox{,}
\end{equation}
where $\alpha_{m, n}$ is a $N \times N$ matrix of effective tight-binding coefficients, and $\phi_{m, n}(k_x, k_y)$ are base functions
\begin{equation}
\phi_{m, n}(k_x, k_y) =  \cos\left( \frac{2 \pi}{a} m k_x\right) \cos\left( \frac{2 \pi}{b} n k_y\right).
\end{equation}
The $\alpha$ matrices are chosen to provide the best fit for the form of the Fermi surface pockets and band velocities at the Fermi level. For the two hole pockets centered at the $\Gamma$ point and for the two electron pockets located at the corners of the BZ, the parameters are as follows (in eV):
%\begin{eqnarray*}
%\alpha^\mathrm{hole}_\mathrm{outer} & = \left  (
%\begin{tabular}{r r r r }
%-0.0651 	   &0.0578	&-0.0048	&0.0074\\
%0.0578	    &0.0857	&-0.0083	&0.0077\\
%-0.0048    &	-0.0083	&0.0115	&0.0069\\
%0.0074	   &0.0077	&0.0069	&-0.0006\\
%\end{tabular}
%\right),\\
%%\end{eqnarray*}
%%\begin{eqnarray*}
%\alpha^\mathrm{hole}_\mathrm{middle}  & = \left  (
%\begin{tabular}{r r r r }
%-0.1831	    &0.0515    &	-0.0124	 &-0.0095\\
%0.0515	    &0.0797	   &0.0248	       &0.0097\\
%-0.0124	   &0.024	   &-0.0005	    &-0.0074\\
%-0.0095	&0.0097	&-0.0074	 &-0.0035\\
%\end{tabular}
%\right),
%\end{eqnarray*}
%\begin{eqnarray*}
%\alpha^\mathrm{el.}_\mathrm{inner}  & \!= \!\left  (
%\begin{tabular}{r r r r }
%0.129&	0.057 \\
%0.057&	-0.07 \\
%\end{tabular}
%\right )\!,
%\alpha^\mathrm{el.}_\mathrm{outer} \!  = \!\left  (
%\begin{tabular}{r r r r }
%0.117 &	0.057 \\
%0.057 &	-0.07 \\
%\end{tabular}
%\right )\!,
%\end{eqnarray*}
%}
\begin{eqnarray*}
\alpha^\mathrm{hole}_\mathrm{outer} & = \left  (
\begin{tabular}{r r r r }
-0.062 & 0.058    & -0.005 & 0.007\\
0.058 & 0.086 & -0.008 & 0.008\\
-0.005 & -0.008 & 0.012 & 0.007\\
0.007 & 0.008 & 0.007 & -0.005\\
\end{tabular}
\right),\\
%\end{eqnarray*}
%\begin{eqnarray*}
\alpha^\mathrm{hole}_\mathrm{middle}  & = \left  (
\begin{tabular}{r r r r }
 -0.192 & 0.052 & -0.012 & -0.009\\
 0.052 & 0.08 & 0.025 & 0.01\\
 -0.012 & 0.025 & -0.001 & -0.007\\
 -0.009 & 0.01 & -0.007 & -0.004\\
\end{tabular}
\right),
\end{eqnarray*}
\begin{eqnarray*}
\alpha^\mathrm{el.}_\mathrm{inner}  & \!= \!\left  (
\begin{tabular}{r r r r }
0.136&	0.057 \\
0.057&	-0.074 \\
\end{tabular}
\right )\!,
\alpha^\mathrm{el.}_\mathrm{outer} \!  = \!\left  (
\begin{tabular}{r r r r }
0.118 &	0.0574 \\
0.057 &	-0.074 \\
\end{tabular}
\right )\!.
\end{eqnarray*}
Although the electron bands are known to have noticeable $k_z$ dispersion we use averaged parameters for the sake of simplicity, thus remaining within a 2D structure.

\begin{figure}[t!]
\centering
\includegraphics[width=1.0\linewidth]{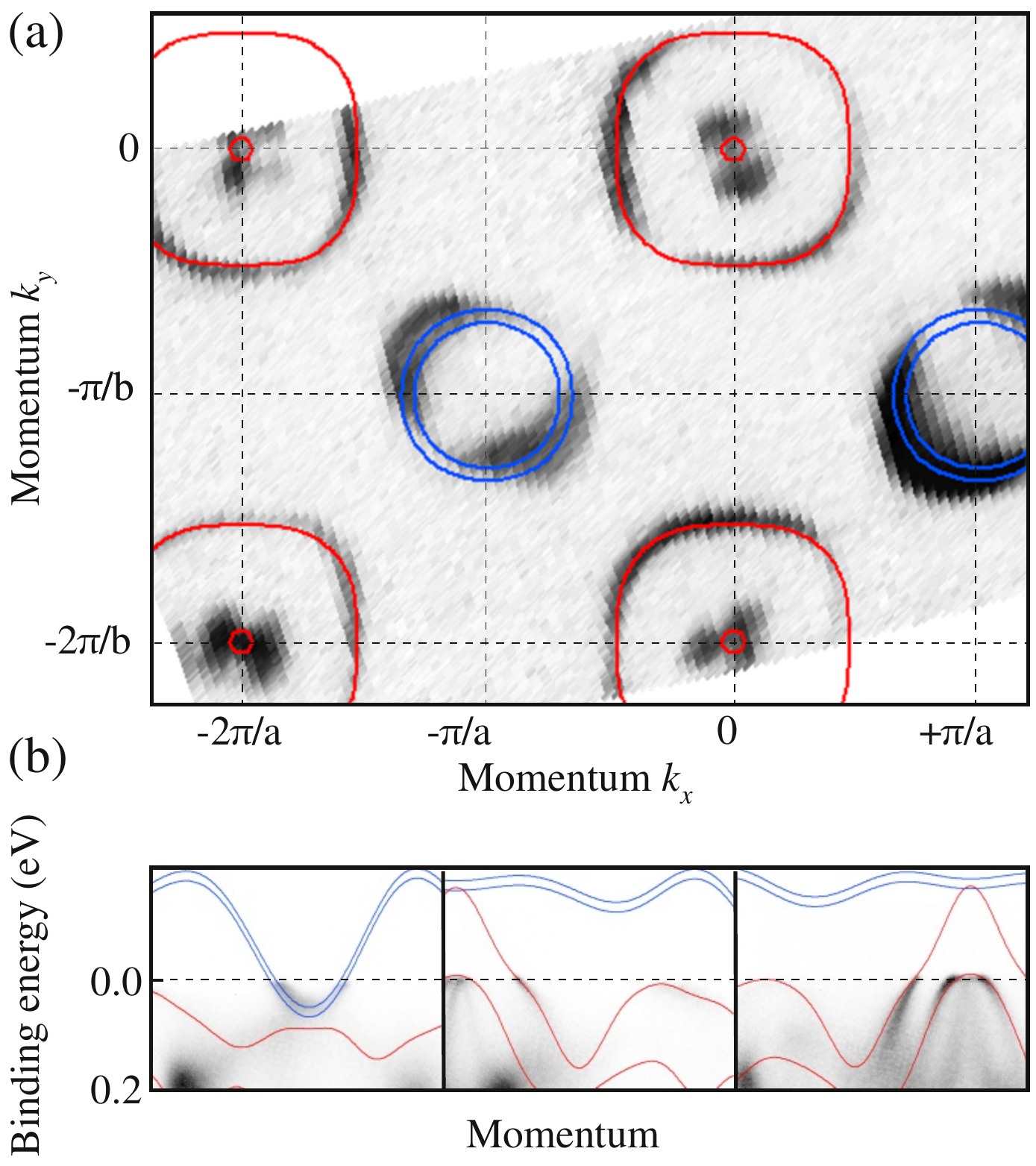}
\caption{(color online) (a) Experimental Fermi surface map with fitting contours superimposed over it. (b) Three general energy--momentum cuts passing through the electron pocket centered at $(-\pi/a;-\pi/b)$ and hole pockets at $(0;0)$ and $(0;-2\pi/b)$.   }
\label{FS_map}
\end{figure}

To demonstrate to which extent this simple model is able to capture the dispersion
of low energy bands, in Fig.\,\ref{FS_map} we plot typical experimental FS maps\cite{borisenko,Stockert,Lankau} with the fitted band dispersion. The lower panel also contains several general energy--momentum cuts, allowing one to compare the experimental and model dispersions for energies close to the Fermi level.

{\it Spin response.}
In the following we proceed with the calculations of the magnetic INS spin response. In the magnetically-disordered state transverse and longitudinal components of the spin susceptibility are identical, and we focus below on the transverse part. The spin response is computed within RPA. Then, the transverse components of the full spin susceptibility $\chi^{i,j}$ are related to transverse components of the bare susceptibility $\chi^{i,j}_{0}$  as
 \begin{equation}
\chi^{i,j}=\chi^{i,j}_{0}+\chi^{i,j^{\prime}}_{0}
 u_{i^\prime,j^\prime}  \chi^{i^{\prime},j},
\label{n_1}
\end{equation}
where $i$ and $j$ are band indices.

Summation over repeated band indices is implied and $u^{i^\prime j^\prime}$ are matrix elements of the interactions. The solution of Eq.(\ref{n_1}) in matrix form is straightforward:
$\hat\chi=\hat\chi_{0}(1-{\hat u} {\hat\chi}_{0})^{-1}$.
The components of the bare spin susceptibility $\hat\chi_{0} =
\chi^{i,j}_0 ({\bf q},{\rm i} \Omega_m)$ are given by usual combinations of normal and anomalous Green's functions
%\begin{widetext}
\begin{eqnarray}
\chi^{ij}_0({\bf p},{\rm i} \Omega_m) & = & - \frac{T}{2N}
\sum_{ {\bf k}, \omega_n} {\rm Tr}
\left[ G^i_{{\bf k + p}}({\rm i} \omega_n + {\rm i} \Omega_m) G^j_{\bf k}({\rm i} \omega_n) \right. \nonumber\\
&& \left.+ F^i_{{\bf k + p}}({\rm i} \omega_n + {\rm i} \Omega_m) F^j_{{\bf k}}({\rm i} \omega_n)\right]\quad,
\label{eq:bare_chi}
\end{eqnarray}
where $G^i_{\bf k}(i\omega_n)=-\frac{i\omega_n+\varepsilon^{i}_{\bf k}}{\omega^2_n+(\varepsilon^{i}_{\bf k})^2+(\Delta_{\bf k
}^{i})^{2}}$ and $F^i_{{\bf k}}(i\omega_n) = \frac{\Delta_{\bf k
}^{i}}{\Omega^2_n+(\varepsilon^{i}_{\bf k})^2+(\Delta_{\bf k
}^{i})^{2}}$. For the superconducting gap function we assume the form, obtained in the ARPES experiments\cite{Symmetry}. In particular, the superconducting gaps on the inner and on the outer hole pockets were found to amount to $\Delta^{h_{inner}} = 6$meV, and $\Delta^{h_{outer}} = 3.4+0.5\left(\cos 4 \phi+ 0.13\cos 8\phi -
0.2\cos 12 \phi\right)$ (in meV), respectively. Here $\phi$ is the angle counted on the hole Fermi surface. For the two electron pockets the gaps were found to be similar in the form  $\Delta^{e_{inner}}=\Delta^{e_{outer}}=3.4+0.5\cos \theta$ (in meV) where $\theta$ is the angle on the electron Fermi surface pockets. Note that a similar angular variation of the superconducting gap on the outer hole pocket was extracted in Scanning Tunneling Microscopy (STM) though with a smaller gap magnitude.\cite{allan}. As ARPES is not sensitive to the phase difference of the gap between electron and hole pockets we consider two possibilities, namely, $s^{+-}$-symmetry of the superconducting gap, where the phase of the superconducting gap changes sign between electron and hole pockets and $s^{++}$ where, despite higher harmonics, the gaps on the electron and hole pockets remains always positive.

In our  numerical calculations we keep all terms
in the matrix equation for the full susceptibility. The interacting part of the Hamiltonian contains four-fermion
interactions with small momentum transfer as well as momentum transfers
around $(\pi,\pi)$.  They include the interactions between electron and hole bands with
momentum transfer around ($\pi$, $\pi$) as well as interactions with small momentum transfer within or between hole pockets and, similarly, for the electron pockets.
For simplicity, we approximate all interactions as angle-independent, i.e., we neglect the angle dependence introduced by dressing the
interactions by coherence factors associated with the hybridization of Fe $d-$orbitals. These coherent factors do play a role in the angular variation of the superconducting gap\cite{saurabh}, but do not substantially modify the positions of the spin resonance~\cite{maier}. For
better convergence of the numerical series we add a small damping of
$\Gamma=3$ meV to the fermionic dispersion in the normal state. This value is consistent with values, extracted from ARPES experiments\cite{borisenko}. We also analyze the influence of the fermionic damping on the spin excitations in the superconducting state below in more detail.

\begin{figure}[t!]
\centering
\includegraphics[width=1.0\linewidth]{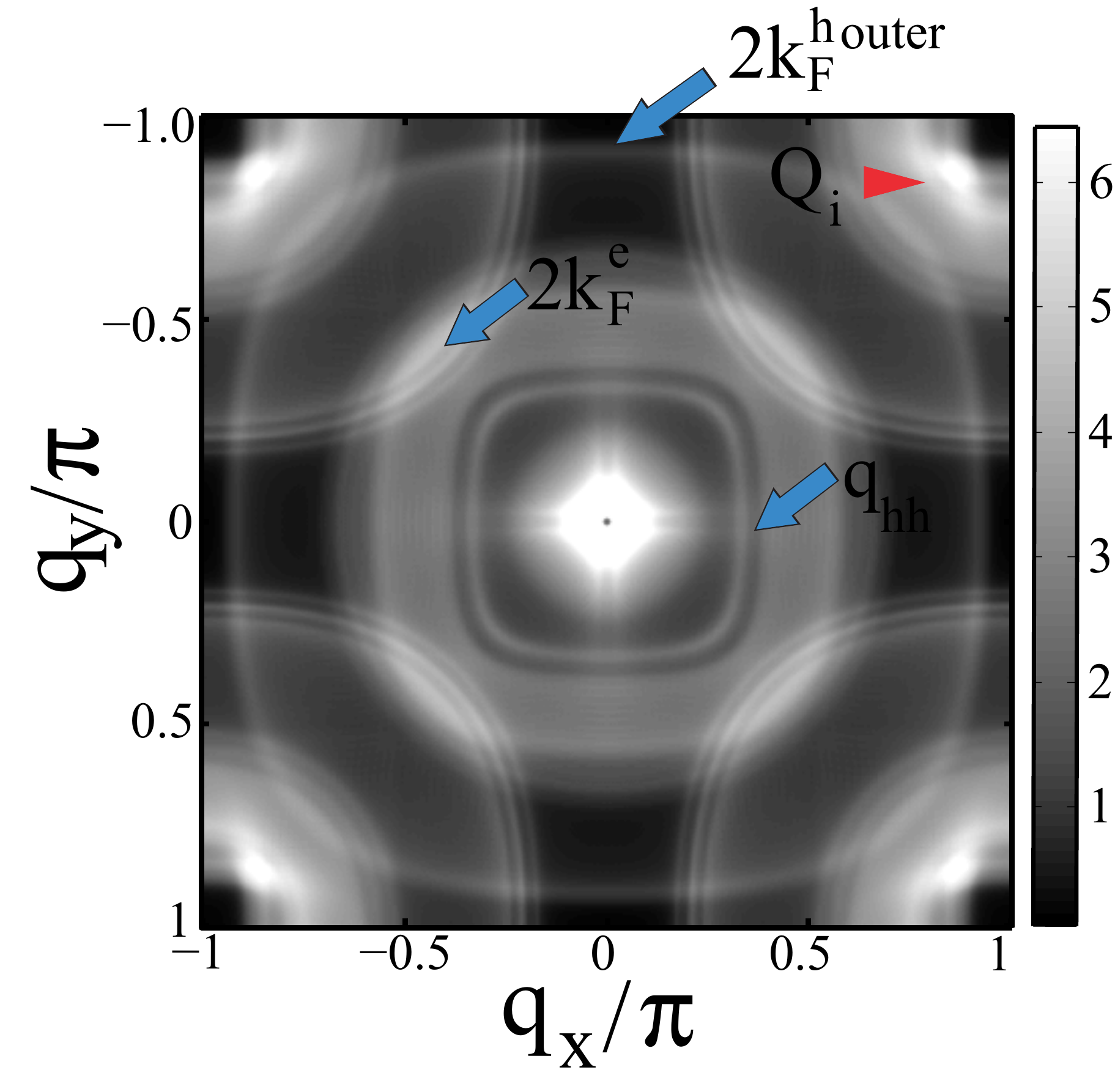}
\caption{(color online) Calculated imaginary part of the bare spin susceptibility in the normal state of LiFeAs as a function of the momentum in the first BZ at $\hbar\Omega$=5meV. }
\label{fig1th}
\end{figure}

\begin{figure}[t!]
\centering
\includegraphics[width=1.0\linewidth]{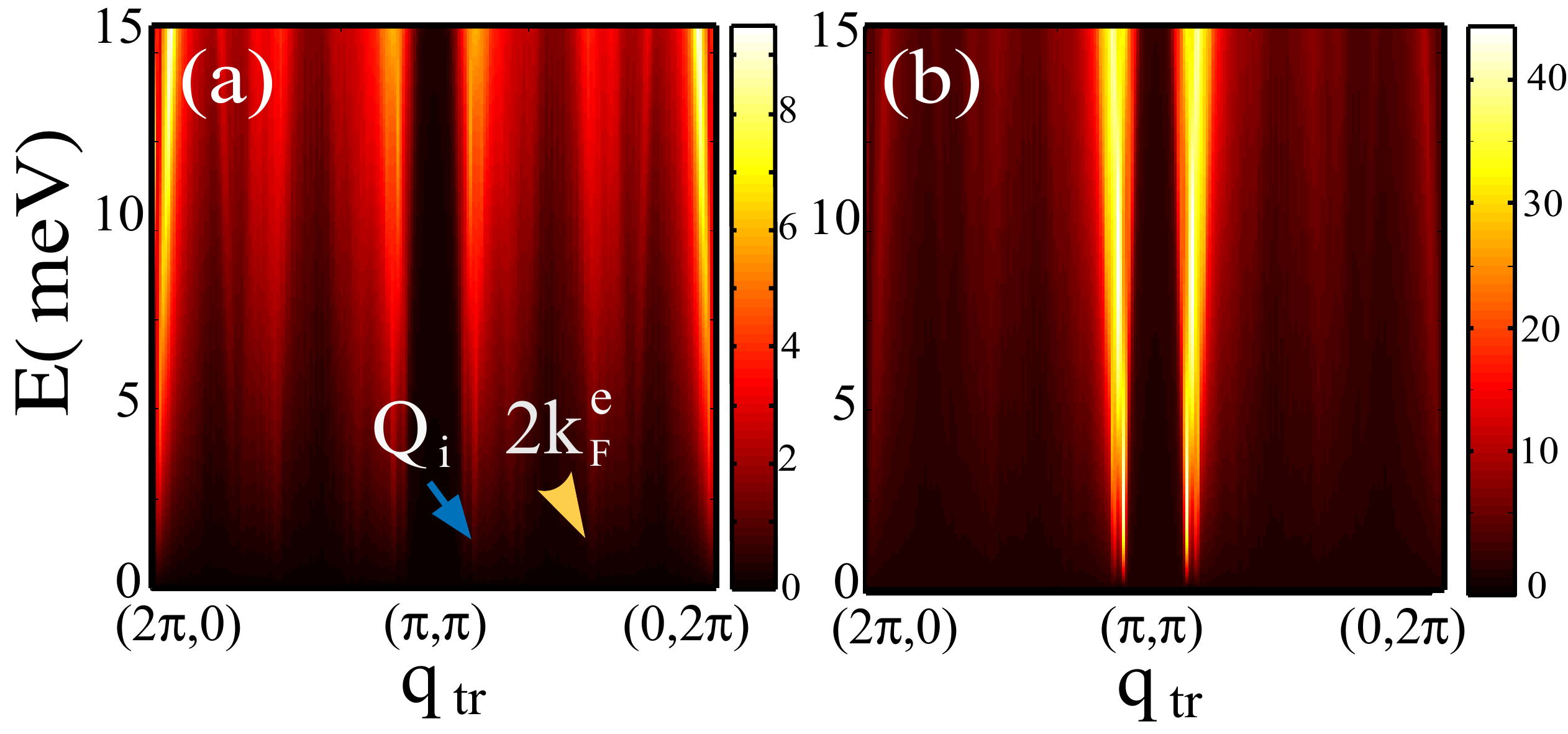}
\caption{(color online) Calculated imaginary part of the bare (a) and the RPA (b) spin response as a function of the transverse momentum and frequency in the normal state of LiFeAs.}
\label{fig2th}
\end{figure}

We start by looking on the bare susceptibility as this quantity directly follows from the fermiology, measured by ARPES. In particular, in Fig.\ref{fig1th} and Fig.\ref{fig2th}(a)  we show its imaginary part as {\bf q} and $\Omega$ maps in the first BZ. As expected, the scattering momenta associated with $2k_F$ intraband processes resemble the original Fermi surfaces  in circular-like structures. Furthermore, Fig.\ref{fig1th} allows for a straightforward identification of the character of the scattering. The intraband scattering and interband scattering between two electron or two hole bands are centered around $(0,0)$, while the interband scattering between the bands of different character are centered around $(\pm \pi, \mp \pi)$ momentum.  By comparing the diameters of the intraband driven scattering circles with the approximate 2$k_F$ values of the corresponding Fermi surfaces one identifies immediately $2k_{F}^{h_{outer}}$ as well as $2k_{F}^{e_{outer}} \sim 2k_{F}^{e_{inner}}$, shown by the arrows. In addition, the bright spot around the ${\bf q}=0$ refers to the scattering within the small inner hole pocket as well as the scattering between the inner and the outer electron pockets.
Further interband scattering processes include the scattering between two hole pockets, denoted by $q_{hh}$ and most importantly, the scattering between the electron pockets and outer hole pocket, shown by ${\bf Q}_i$. Note that the scattering between the inner hole pocket and the electron pockets occurs also at a similar momentum, but its intensity is much smaller already in the bare susceptibility due to limited phase space available for scattering. Within the RPA the intensity of these excitations is further suppressed as compared to the scattering between the outer hole pocket and electron pockets. Therefore, we can safely conclude that the excitations at {\bf Q}$_i$ arise due to scattering between the large outer hole pocket and the two electron pockets. Note, our analysis does not include the matrix elements, originating from the transformations from the orbital basis to the band ones. Its inclusion usually strengthens the transverse  scattering, {\bf q}$_{tr}=(q,2\pi-q)$, over the longitudinal ones {\bf q}$_{lg}=(q,q)$\cite{park}. This would make the intensity, shown in Fig.2, look more anisotropic, but it will not change the position of the peaks. Note that our calculations within a three-orbital model are consistent with this observation.

These scattering wave vectors are also visible along the transverse direction, $q_{tr}=(q,2\pi-q)$, shown in Fig.\ref{fig2th}(a). The scattering within the small inner hole pocket at small momentum {\bf q} as well as the scattering between the electron pockets and outer large hole pocket at the wave vector {\bf Q}$_i$ are most pronounced. One can further identify the intraband scattering within the electron bands at $2k_{F}^{e}$. Note also that on the energy scale from 0 to 15 meV the dispersion of these excitations is almost vertical, which is caused  by the relatively large Fermi velocity of the involved bands.
By comparison with the INS we find that the scattering momentum {\bf Q}$_i \approx (0.86,1.14)\pi$, associated with the scattering between the outer hole pocket and the two electron pockets, matches precisely with the experimentally observed incommensurate momentum. As our band structure results from the fit to the ARPES band structure, we conclude that the incommensurate momentum seen in the INS refers to the scattering between the electron and the outer hole Fermi surfaces. This can be further supported by the fact that the incommensurate magnetic excitations, found in INS, do not indicate a strong $z$-dispersion. We recall that the small inner hole pocket around the $\Gamma-$point of the BZ has strongly three-dimensional character which we ignored at present. However, if taken into account, it should produce a strong dispersion of the incommensurate magnetic excitations along the $q_z$ momentum, which is not the case. Therefore, the scattering between the outer hole pocket and the two electron pockets is, most likely, responsible for the INS intensity at the wavevector {\bf Q}$_i \approx (0.86,1.14)\pi$, which differs from the proposal made in Ref.\cite{wang_dai_new} where these incommensurate peaks were attributed to the scattering between the small inner hole pocket and two electron pockets.

To proceed further, we compute the total RPA susceptibility by including the interactions. Note, however, that  part of these interaction parameters contributes already to the renormalization of the bands, which was used to obtain the tight-binding model. As the interaction values are not fully known we took them to be the same, {\it i.e.} $u=u_{interband}=u_{interband}=0.78 \alpha^{hole}_{outer}(1,2)$ except for the scattering between the electron pockets, which we consider to be small $u_{ee}=0.1u$. The magnitudes of the interactions were also chosen such that the system remains in the paramagnetic phase. Fig.\ref{fig2th}(b) shows the results for the Im$\chi_{RPA}$, displayed as $q_{tr}$ and $E=\hbar\Omega$ map. Observe that in comparison to the bare susceptibility the incommensurate excitations due to the scattering between electron and hole bands are enhanced.
At the same time, we find that the excitations at small {\bf q} are much less intense as compared to the bare $\chi_0$. To understand the origin of its suppression at finite frequencies recall that the real part of the bare intraband susceptibility falls off as $1/\Omega$ which indicates that within RPA there is no source for the enhancement of these small {\bf q} excitations. This explains why the total susceptibility shows stronger enhancement only for the wavevector {\bf Q}$_i$ and not for the ${\bf q} \sim 0 $ momentum.

\begin{figure}[t!]
\centering
\includegraphics[width=1.0\linewidth]{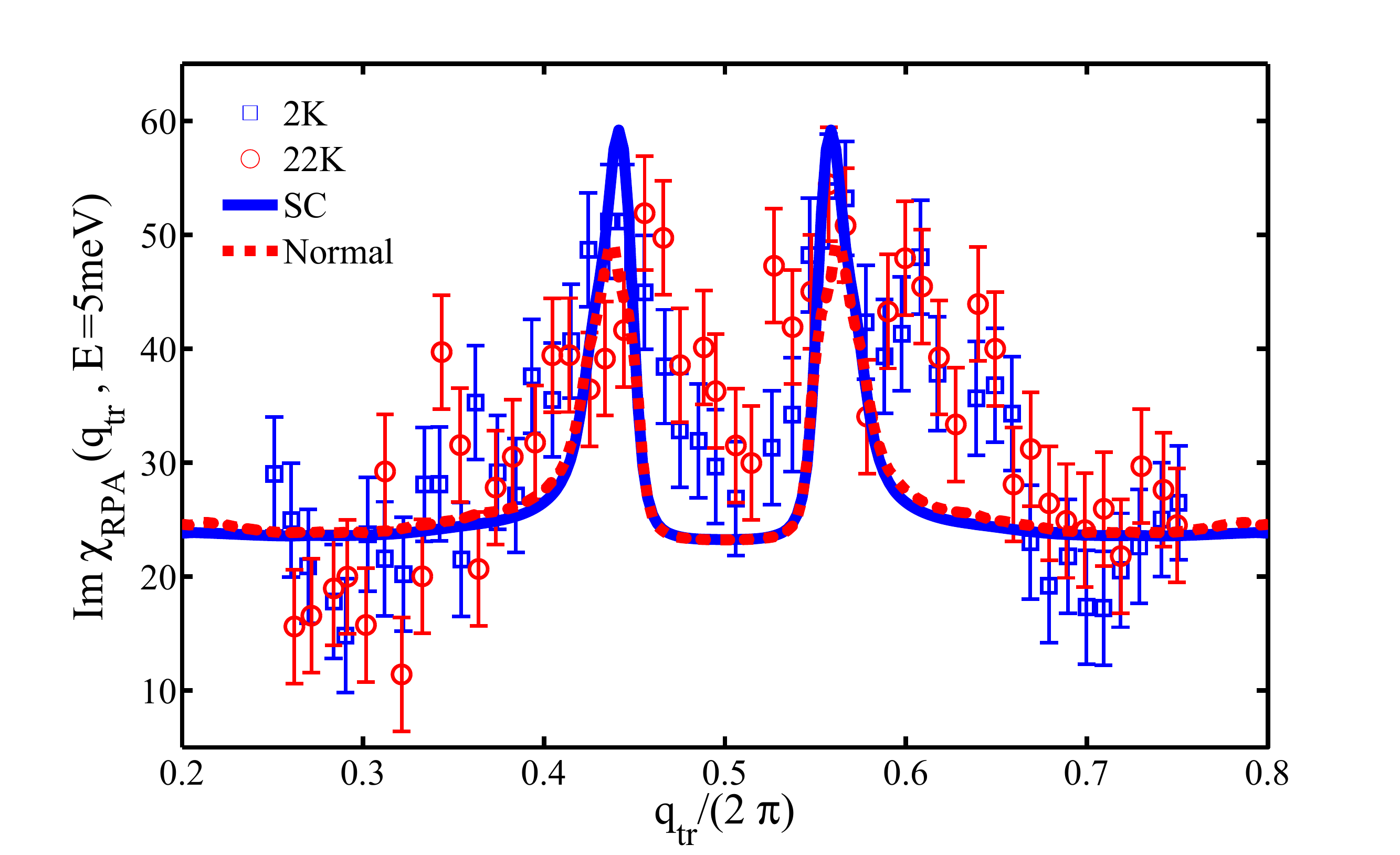}
\caption{(color online) Calculated momentum dependence of the the imaginary part of the RPA  spin response at $\hbar \Omega=5$meV
in the normal and suerconducting $s^{+-}$-wave states, respectively. The symbols refer to the experimental data, taken from Ref.\cite{qureshi}.}
\label{fig3th}
\end{figure}
A comparison to the experimental INS data in the normal state is shown in Fig.\ref{fig3th} where we display the imaginary part of the total RPA susceptibility for $\hbar \Omega=5$meV as a function of the transverse momentum. The pronounced peaks at ${\bf Q}_i$ due to the scattering between the outer hole and two electron pockets agree with those found experimentally\cite{qureshi}. Nevertheless, one should mention that the structure of the peaks is symmetric in the calculations, while in INS there is an additional shoulder for larger $q_{tr}$. A weak $z$-dispersion or some other scattering paths may cause this behavior.

In the next step, we move to the superconducting state and compute the spin excitations for various symmetries of the superconducting order parameters. The most interesting question is whether any information can be extracted about the phase structure of the gap with respect to the relative phase difference between electron and hole pockets as well as between inner and outer hole pockets. We remind that in contrast to the angular dependence of the gap, the relative phase structure cannot be directly probed by ARPES. In particular, we considered two different situations. The first one, which we name $s^{+-}$, refers to the phase of the superconducting gap on the hole pockets being opposite to the phase of the gap on the electron pockets. In the other case, the so-called $s^{++}$, the overall phase of the order parameter is the same for the electron and the hole pockets.

\begin{figure}[t!]
\centering
\includegraphics[width=1.0\linewidth]{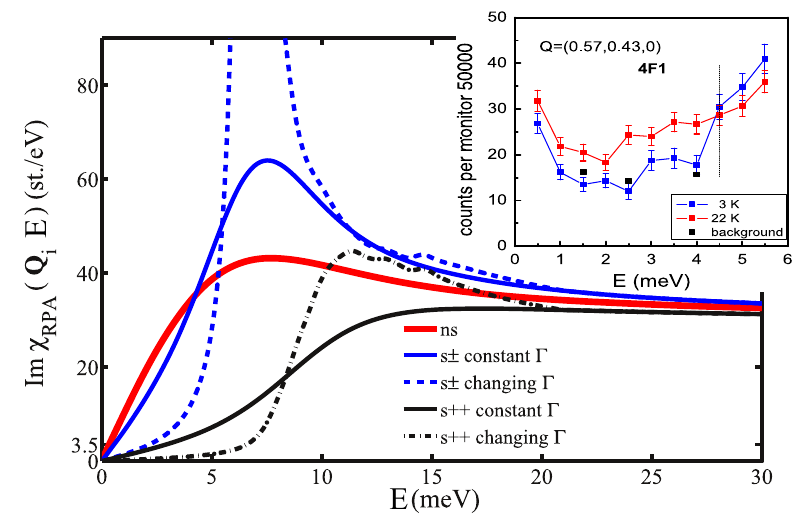}
\caption{ Calculated frequency dependence of  the imaginary part of the RPA  spin response at the wavevector {\bf Q}$_i$ in the normal state and the superconducting state for two different symmetries of the superconducting gap. Difference curves for the $s^{++}$ and $s^{+-}$-wave scenarios refer to the constant fermionic damping of 3 meV and the one frequency dependent, as described in the text. The inset represent the experimental data, taken from Ref.\cite{qureshi}. Observe that the upturn behavior of the experimental curves at energies smaller than 1.5 meV is due to elastic scattering contribution.}
\label{fig4th}
\end{figure}

In Fig.\ref{fig3th} we show the behavior of spin excitations for the $s^{+-}$-wave symmetry together with the normal state results and the experimental data for $\hbar \Omega=5$meV. Observe that the renormalization of the spin excitations in the $s^{+-}-$wave channel is present but relatively weak in the sense that the excitations are only slightly enhanced  with respect to the normal state. The reason for this moderate renormalization is the relatively strong angular variation of the superconducting gap on the outer hole pocket and on the electron pockets\cite{maiti_res}. The angular-dependent gap washes out the strong enhancement of Im$\chi_0$ at 2$\Delta_0$ which is a prerequisite for the sharp resonance. In LiFeAs the enhancement of Im$\chi_0$ in the superconducting state with respect to its normal state value is more gradual than in the other FeAs superconductors. In addition the gaps on the electron and on the outer hole pockets contributing mostly at this ${\bf Q}_i$ are relatively small. Taken together with the constant value of the fermionic damping of the order of $\sim 3$meV, these factors render the enhancement of the spin excitations in the $s^{+-}$-wave scenario relatively weak.
In other words in LiFeAs there is no true spin resonance below the continuum of particle-hole excitations but an enhancement of the continuum itself due to coherence factors associated with the phase structure of the $s^{+-}$-wave superconducting gap. This also makes it difficult to distinguish the other scenarios. In particular, although an $s^{+-}$-wave gap structure produces spin excitations which agree with available experimental data, the results with an $s^{++}$-wave symmetry cannot be ruled out based on the measurements at a given frequency. The difference between two symmetries of the order parameters becomes more apparent in Fig.\ref{fig4th} where we plot Im $\chi_{RPA} = \sum_{i,j} \chi^{i,j}$ in the superconducting and in the normal state, as a function of frequency ($\hbar \Omega$) at the wavevector {\bf Q}$_i$. Notice first that the normal state curve shows a characteristic single relaxor form of the overdamped paramagnons, centered around 7.5 meV. It agrees qualitatively with the INS data, see Fig.3(b) in Ref.\cite{qureshi}.

In the superconducting state the results for the s$^{++}$ and the s$^{+-}$-wave superconducting gaps depend sensitively upon the assumption made for the fermionic damping. For constant damping the $s^{+-}$ is the only symmetry which qualitatively agrees with the experimental data of Ref.\cite{qureshi}, shown in the inset of Fig.\ref{fig4th}. In particular, one finds that the intensity of the spin excitations is suppressed  with respect to its normal state values up to energies of about 4.2 meV and is then slightly enhanced for higher energies. However, as mentioned above this enhancement is not a true exciton but rather an enhancement of the particle-hole continuum due to the sign change of the superconducting gap. At the same time, for the $s^{++}$-wave symmetry the difference between the superconducting and normal states remains always  negative for energies up to 30 meV for constant damping. Here, the spin excitations are suppressed as the superconducting gap does not change sign at this particular {\bf Q}$_i$. We note, however, that manipulation of the fermionic damping improves the situation. Following Ref.\cite{kontani} we modeled the fermionic damping in the superconducting state as $\Gamma \sim 0$ for  $0<\hbar \Omega<3\Delta_{ave}$, $\Gamma=\Gamma_{ns}$ for an $\hbar \Omega>4\Delta_{ave}$, and increasing linearly for $3\Delta_{ave}<\hbar \Omega<4\Delta_{ave}$. We varied the value for $\Delta_{ave}$ between 3 and 6 meV to find the best-case scenario for the $s^{++}$-wave symmetry regarding the enhancement in the superconducting state. The result for $\Delta_{ave}\sim 5$meV is shown in Fig.\ref{fig4th}. For the chosen damping parameters, the s$^{++}$-wave symmetry exhibits
scattering enhancement in the superconducting state while
a true spin exciton at the energy of about 6meV
appears for the s$^{+-}$-wave symmetry. However, the enhancement for s$^{++}$ occurs at higher energies
than in the experiment which clearly indicates intensity gain in the
superconducting phase for energies between 5 and 10 meV.
The agreement with s$^{++}$-wave symmetry  can be improved by reducing the
$\Delta_{ave}$ value, but the magnitude of the enhancement with
respect to the normal state then also becomes smaller.
At this point, we can only conclude that experimental results do not show a true spin resonance mode,
frequently taken as characteristic for the s$^{+-}$-superconductor,
but a redistribution of the particle-hole continuum in the presence of
superconductivity. This renders the definite conclusion on
the phase structure of the superconducting gap rather
difficult.

To conclude, we employ the ARPES data for LiFeAs to make a connection to the INS response. A comparison to the INS data shows that the incommensurate magnetic scattering intensity originates from the scattering between the electron pockets, centered around the $(\pi,\pi)$ point of the BZ and the large two-dimensional hole pocket, centered around the $\Gamma$-point of the BZ. This points towards an internal consistency between the FS topology, measured by ARPES, and INS results. We also find that the renormalization of the neutron intensity in LiFeAs in the superconducting state can be understood  in terms of a rearrangement of the particle-hole continuum which is rather weak for any phase structure of the superconducting gap between electron and hole pockets. Further studies are necessary to elucidate the nature of the Cooper-pairing in this compound.

We thanks J. van den Brink, A.V. Chubukov, and P.J. Hirschfeld for fruitful discussions.
IE acknowledges financial support of the Merkur Foundation, German Academic Exchange Service (DAAD PPP USA No. 50750339). JK acknowledges support from the Studienstiftung des deutschen Volkes, IMPRS Dynamical Processes  in Atoms, Molecules and Solids, and DFG within GRK1621. VZ and SB acknowledge support by DFG grants ZA 654/1-1 and BO 1912/3-1 (SPP1458). NQ and MB were
supported by the DFG through SFB 608. The work is supported by DFG grants BE1749/13  (SPP1458) and the Graduiertenkolleg GRK1621 and by the Dresden Platform for Superconductivity and Magnetism.

\end{document}